\documentclass[a4paper,11pt]{article}
\usepackage{pos}

\numberwithin{equation}{section}
\usepackage{float}

\usepackage{lipsum}

\title{Correlated Dirac eigenvalues around the transition temperature on $N_{\tau}=8$ lattices}
\ShortTitle{Correlated Dirac eigenvalues}

\author[1]{Heng-Tong Ding}
\author*[1]{Wei-Ping Huang}
\author[1]{Min Lin}
\author[2]{Swagato Mukherjee}
\author[2]{Peter Petreczky}
\author[3]{Yu Zhang}

\affiliation[1]{Key Laboratory of Quark \& Lepton Physics (MOE) and Institute of Particle Physics,\\
  Central China Normal University, Wuhan 430079, China}

\affiliation[2]{Physics Department, Brookhaven National Laboratory, Upton, NY 11973, USA}

\affiliation[3]{RIKEN Center for Computational Science, 7-1-26	Minatojima-minami-machi, Chuo-ku, Kobe, Hyogo 650-0047, Japan}

\emailAdd{hengtong.ding@mail.ccnu.edu.cn}
\emailAdd{huangweiping@mails.ccnu.edu.cn}
\emailAdd{linmin@mails.ccnu.edu.cn}
\emailAdd{swagato@bnl.gov}
\emailAdd{petreczk@bnl.gov}
\emailAdd{yu.zhang.ey@riken.jp}

\abstract{
We investigate the criticality of chiral phase transition manifested in the first and second order derivatives of Dirac eigenvalue spectrum with respect to light quark mass in (2+1)-flavor lattice QCD. 
Simulations are performed at temperatures from about 137 MeV to 176 MeV on $N_{\tau}=8$ lattices using the highly improved staggered quarks and the tree-level improved Symanzik gauge action. The strange quark mass is fixed to its physical value $m_s^{\text{phy}}$ and the light quark mass is set to $m_s^{\text{phy}}/40$ which corresponds to a Goldstone pion mass $m_{\pi}=110$ MeV. 
We find that in contrast to the case at $T\simeq 205$ MeV
$m_l^{-1} \partial \rho(\lambda, m_l)/\partial m_l$ is no longer equal to $\partial ^2\rho(\lambda, m_l)/\partial m_l^2$ and $\partial ^2\rho(\lambda, m_l)/\partial m_l^2$ even becomes negative at certain low temperatures. This means that as temperature getting closer to $T_c$ $\rho(\lambda, m_l)$ is no longer  proportional
to $m_l^2$ and thus dilute instanton gas approximation is 
not valid for these temperatures.
We demonstrate the temperature dependence can be factored out in $\partial \rho(\lambda, m_l)/ \partial m_l$ and $\partial^2 \rho(\lambda, m_l)/ \partial m_l^2$ at $T \in [137, 153]$ MeV, and then we propose a feasible method to estimate the power $c$ given $\rho \propto m_l^{c}$.
}

\FullConference{%
 The 38th International Symposium on Lattice Field Theory, LATTICE2021
  26th-30th July, 2021
  Zoom/Gather@Massachusetts Institute of Technology
}

\begin{document}

\maketitle

\section{Introduction}

In the classical and chiral limit the theory of (2+1)-flavor Quantum chromodynamics (QCD) has the $U(1)_V \otimes U(1)_A \otimes SU(2)_L \otimes SU(2)_R$ symmetry. Due to the spontaneous chiral symmetry breaking and the quantum anomaly, this large group of symmetries breaks down to a smaller subgroup $SU(2)_V\otimes U(1)_V$.
At the physical point, where the quark masses are set to their physical values, the QCD medium undergoes a continuous crossover from the hadronic phase to the quark-gluon plasma (QGP) phase at a pseudo-critical temperature $T_{pc} \simeq 156$ MeV \cite{HotQCD:2018pds, Borsanyi:2020fev}.
While in the chiral limit of light quark mass, the chiral symmetry restores through a second order phase transition belonging to 3-dimensional $O(4)$ universality class \cite{Pisarski:1983ms, Ding:2020xlj} at a estimated chiral critical temperature $T_c=132^{+3}_{-6}$ MeV \cite{HotQCD:2019xnw}.

Detailed investigations on the restoration of the chiral and axial $U(1)_A$ symmetry in lattice QCD can be made through the study of the differences between various chiral susceptibilities $\chi_{H}$ \cite{Bazavov2012}. These susceptibilities are defined as integrated two-point correlation functions of the quark bilinear $J_H(x) = \bar{q}(x)\Gamma_H q(x)$ for various meson channels $H=\sigma, \delta, \pi ~\mathrm{and}~\eta$, i.e., $\chi_{H}=\int d^{4} x\langle J_{H}(x) J_{H}^{\dagger}(0)\rangle$.
The restorations of the chiral and the axial $U(1)_A$ symmetry lead to $\chi_{\pi}-\chi_{\sigma}=\chi_{\delta}-\chi_{\eta}=0$ and $\chi_{\pi}-\chi_{\delta}=\chi_{\sigma}-\chi_{\eta}=0$, respectively \cite{Bazavov2012}.
Further, by using the technique of spectrum decomposition of the quark propagator \cite{Smilga1993} chiral observables like chiral condensate and above-mentioned chiral susceptibilities can be related to the Dirac eigenvalue spectrum $\rho(\lambda, m_l)$ and its first derivative with respect to quark mass $\partial \rho(\lambda, m_l) / \partial m_{l}$. It has been found very recently that the n-th order quark mass derivative of $\rho(\lambda, m_l)$, i.e., $\partial ^n\rho(\lambda, m_l) / \partial m_{l}^n$, is related to correlations among Dirac eigenvalues on the each gauge ensemble~\cite{Ding:2020xlj}.

It has been demonstrated in Ref.~\cite{Ding:2020xlj} that at $T\simeq 1.6 T_c\approx205$ MeV the axial U(1) anomaly remains manifested in $\chi_{\pi}-\chi_{\delta}$ as well as the disconnected chiral susceptibility $\chi_{disc}$, and $\rho(\lambda, m_l)$ in the infrared region is proportional to $m_l^2$ and develops a peaked structure by studying the first, second and third derivatives of $\rho(\lambda,m_l)$ with respect to $m_l$. On the other hand, in the vicinity of the phase transition temperature the critical behaviors of the chiral order parameter $M(t, h)$ (the chiral condensate) and its susceptibility $\chi_M(t, h)$ can be described by the so-called magnetic equation of state (MEOS) \cite{Ejiri:2009ac}, which is controlled by the scaling functions that are characteristic for different universality class of the phase transition. $M$, $\chi_M$ and the first order quark mass derivative of $\chi_M$ are expressed in terms of scaling functions as follows
\begin{align}
M(t, h)
&=h^{1 / \delta} f_{G}(z)~,
\\
\chi_{M}(t, h)
&=\frac{\partial M}{\partial H}
=\frac{1}{h_{0}} h^{1 / \delta-1} f_{\chi}(z)~,
\\
\label{fpp}
\frac{\partial \chi_M}{\partial H} 
&= \frac{1}{h_{0}^{2}} \frac{1}{\delta} h^{1 / \delta-2} f_{\mathrm{pp}}(z)~.
\end{align}
where the scaling variable $z=t / h^{1 / \beta \delta}$ with $\beta$ and $\delta$ the critical exponents. Here $t$ and $h$ are reduced temperature and symmetry-breaking field respectively, and $h_0$ is a non-universal parameter. $f_G(z)$, $f_{\chi}(z)$ and $f_{\mathrm{pp}}(z)$ are the corresponding scaling functions.
Current studies of the scaling behaviors are mainly based on the chiral condensate and its susceptibility \cite{HotQCD:2019xnw, Ejiri:2009ac}.

It thus would be interesting to study how the criticality of chiral phase transition is manifested in $\rho(\lambda, m_l)$ and its quark mass derivatives connected with the correlation among Dirac eigenvalue spectrum. In this work we extend the study in Ref.~\cite{Ding:2020xlj}, where only a single temperature $T\simeq 205$ MeV was studied, to a temperature window $T\in$[137,176] MeV in the vicinity of $T_c(N_\tau=8)\approx143$ MeV~\cite{HotQCD:2019xnw}. The current study is based on the $N_f=2+1$ lattice QCD simulations using Highly Improved staggered fermions (HISQ) on $N_\tau=8$ lattices with $m_\pi=110$ MeV.

The proceeding is oragnized as follows.
In section 2 the recently proposed relation between $\partial ^n\rho(\lambda, m_l) / \partial m_{l}^n$ and correlations among Dirac eigenvalue spectrum on each gauge ensembles is briefly reviewed. Based on this relation we can directly compute $\partial ^n\rho(\lambda, m_l) / \partial m_{l}^n$ on the lattice.
In section 3 we show details of our lattice setup.
In section 4 we present the temperature dependences of various chiral observables as well as $\rho$ and its first and second quark mass derivatives.
In section 5 we demonstrate the temperature dependence can be factored out in $\partial \rho(\lambda, m_l)/ \partial m_l$ and $\partial^2 \rho(\lambda, m_l)/ \partial m_l^2$ at $T \in [137,153]$ MeV, and then propose a feasible method to estimate the power $c$ if assuming $\rho \propto m_l^{c}$.

\section{Correlations among Dirac eigenvalues and mass derivatives of Dirac eigenvalue spectrum}

It has been proposed that the n-th order derivative of Dirac eigenvalue spectrum with respect to quark mass $\partial^n\rho(\lambda, m_l)/\partial m_l^n$ can be related to the (n+1)-point correlation function $C_{n+1}$ among the massless Dirac eigenvalues \cite{Ding:2020xlj}.
Dirac eigenvalue spectrum $\rho(\lambda, m_l)$ is defined as the functional integration of $\rho_U(\lambda)$ over the gauge fields, and for (2+1)-flavor QCD it is expressed as
\begin{align}\label{rho}
\rho\left(\lambda, m_{l}\right)
=\frac{T}{V Z[\mathcal{U}]}
\int \mathcal{D}[\mathcal{U}] e^{-S_{G}[\mathcal{U}]}
\operatorname{det}\left[ D \mkern -11 mu / ~[\mathcal{U}]+m_{s}\right] 
\times
\left(\operatorname{det}\left[ D \mkern -11 mu / ~[\mathcal{U}]+m_{l}\right]\right)^{2} 
\rho_{U}(\lambda) ~.
\end{align}
Here $\rho_U(\lambda) = \sum_{j} \delta (\lambda-\lambda_{j})$ and $\lambda_j$ are the eigenvalues of  the massless Dirac matrix $D \mkern -11 mu / ~[\mathcal{U}]$ in a given gauge field $\mathcal{U}$. As shown in Eq.(\ref{rho}) the light quark mass dependence of $\rho(\lambda, m_l)$ is introduced by the functional integration of $\rho_U(\lambda)$ over the gauge fields.
By expressing the fermion determinant as \cite{Ding:2020xlj}
\begin{align}
\label{c1}
\operatorname{det}[ D\mkern -11 mu /~[\mathcal{U}]+ m_{l}]
=\prod_{j} (+\mathrm{i} \lambda_{j}+m_{l})(-\mathrm{i} \lambda_{j}+m_{l})
=\exp \left\{
\int_{0}^{\infty} \mathrm{d} \lambda \rho_{U}(\lambda) \ln \left[\lambda^{2}+m_{l}^{2}\right]
\right\}~,
\end{align}
and then substituting Eq.(\ref{c1}) in Eq.(\ref{rho}), we can directly compute the derivatives of Eq.(\ref{rho}) with respect to $m_l$ \cite{Ding:2020xlj}
\begin{equation}
\begin{aligned}
\frac{\partial \rho}{\partial m_{l}}
=\frac{T}{V} \int_{0}^{\infty} \mathrm{d} \lambda_{2} \frac{4 m_{l} C_{2}\left(\lambda, \lambda_{2} ; m_{l}\right)}{\lambda_{2}^{2}+m_{l}^{2}} ~,
\end{aligned}
\end{equation}
\begin{equation}
\begin{aligned}\label{par2expression}
\frac{\partial^{2} \rho}{\partial m_{l}^{2}}
=\frac{T}{V}\int_{0}^{\infty} \mathrm{d} \lambda_{2} \frac{4(\lambda_{2}^{2}-m_{l}^{2}) C_{2}\left(\lambda, \lambda_{2} ; m_{l}\right)}{(\lambda_{2}^{2}+m_{l}^{2})^{2}} 
+\int_{0}^{\infty} \mathrm{d} \lambda_{2} \mathrm{~d} \lambda_{3} \frac{(4 m_{l})^{2} C_{3}\left(\lambda, \lambda_{2}, \lambda_{3} ; m_{l}\right)}{(\lambda_{2}^{2}+m_{l}^{2})(\lambda_{3}^{2}+m_{l}^{2})}~.
\end{aligned}
\end{equation}
Here $C_2$ and $C_3$ are the 2-point and 3-point correlation functions among $\rho_U(\lambda)$, respectively
\begin{equation}
\begin{aligned}
C_{2}\left(\lambda, \lambda_{2} ; m_{l}\right)=\left\langle\rho_{U}(\lambda) \rho_{U}\left(\lambda_{2}\right)\right\rangle-\left\langle\rho_{U}(\lambda)\right\rangle\left\langle\rho_{U}\left(\lambda_{2}\right)\right\rangle~,
\end{aligned}
\end{equation}
\begin{equation}
\begin{aligned}
C_{3}\left(\lambda, \lambda_{2}, \lambda_{3} ; m_{l}\right)
=&\left\langle\rho_{U}(\lambda) \rho_{U}\left(\lambda_{2}\right) \rho_{U}\left(\lambda_{3}\right)\right\rangle
-\left\langle\rho_{U}(\lambda)\right\rangle\left\langle\rho_{U}\left(\lambda_{2}\right) \rho_{U}\left(\lambda_{3}\right)\right\rangle 
\\ 
-&\left\langle\rho_{U}\left(\lambda_{2}\right)\right\rangle\left\langle\rho_{U}(\lambda) \rho_{U}\left(\lambda_{3}\right)\right\rangle
-\left\langle\rho_{U}\left(\lambda_{3}\right)\right\rangle\left\langle\rho_{U}(\lambda) \rho_{U}\left(\lambda_{2}\right)\right\rangle 
\\ 
+&2\left\langle\rho_{U}(\lambda)\right\rangle\left\langle\rho_{U}\left(\lambda_{2}\right)\right\rangle\left\langle\rho_{U}\left(\lambda_{3}\right)\right\rangle~.
\end{aligned}
\end{equation}

\section{Lattice setup}

 (2+1)-flavor lattice QCD simulations are performed at temperatures from about 137 MeV to 176 MeV on $N_{\tau}=8$ lattices using the highly improved staggered quarks and the tree-level improved Symanzik gauge action. 
In our simulations the strange quark mass is fixed to its physical value $m_s^{\text{phy}}$, and the light quark mass is set to $m_s^{\text{phy}}/40$ which correspond to a Goldstone pion mass $m_{\pi}=110$ MeV.
$\rho_U(\lambda)$ over the entire range of $\lambda$ as well as their $n$-point correlations $C_n(\lambda_1,\cdots,\lambda_n;m_l)$ with $n\leq3$ were computed using the Chebyshev filtering technique combined with the stochastic estimate method \cite{Ding:2020xlj, Zhang2020} on about $10^3$ configurations, where each configuration is separated by 20 time units.
Orders of the Chebyshev polynomials were chosen to be $1.6 \times10^4$ and 20 Gaussian stochastic sources were used. 
Direct measurements of the chiral condensate, $\chi_{\pi}-\chi_{\delta}$, $\chi_{\mathrm{disc}}$ and $\partial \chi_{\mathrm{disc}} / \partial m_l$ on each of these ensembles are also performed through the inversion of the light fermion matrix using 20 Gaussian random sources.

\section{Quark mass derivatives of Dirac eigenvalue spectrum and reproduction of chiral observables}

By using the spectrum decomposition of the quark propagator, a two-flavor light quark chiral condensate $\langle\bar{\psi} \psi\rangle_{l}$ and the difference of susceptibilities of $\pi$ and $\delta$ channels can be expressed in the form of $\rho(\lambda,m_l)$ as
\begin{equation}
\label{pbp_l}
\langle\bar{\psi} \psi\rangle_{l}
=\int_{0}^{\infty} \mathrm{d} \lambda~
\frac{4 m_{l} \cdot \rho\left(\lambda, m_{l}\right)}{\lambda^{2}+m_{l}^{2}} ~,
\end{equation}
\begin{equation}\label{chi_pi-chi_delta}
\chi_{\pi}-\chi_{\delta}
=\int_{0}^{\infty} \mathrm{d} \lambda \frac{8 m_{l}^{2} \rho(\lambda, m_l )}{(\lambda^{2}+m_{l}^{2})^{2}}~.
\end{equation}
Here $m_d=m_u\equiv m_l$. And the two-flavor light quark disconnected susceptibility $\chi_{\mathrm{disc}}$ can be related to $\partial \rho(\lambda,m_l ) / \partial m_{l}$ through
\begin{equation}
\begin{aligned}\label{chi_disc}
\chi_{\mathrm{disc}}=\int_{0}^{\infty} \mathrm{d} \lambda~
\frac{4 m_{l} \cdot \partial \rho\left(\lambda, m_{l}\right) / \partial m_{l}}{\lambda^{2}+m_{l}^{2}}~.
\end{aligned}
\end{equation}

We can further express the mass derivative of the disconnected susceptibility $\partial \chi_{\mathrm{disc}}/\partial m_l$ in terms of first and second order mass derivatives of $\rho(\lambda,m_l)$ as follows
\begin{equation}\label{mass_chi_rho}
\frac{\partial \chi_{\text {disc }}}{\partial m_{l}}
=
\int_{0}^{\infty} \mathrm{d} \lambda \frac{4 m_{l} \partial^{2} \rho / \partial m_{l}^{2}}{\lambda^{2}+m_{l}^{2}}
+
\int_{0}^{\infty} \mathrm{d} \lambda \frac{4(\lambda^{2}-m_{l}^{2}) \partial \rho / \partial m_{l}}{(\lambda^{2}+m_{l}^{2})^{2}}
\equiv \chi_2 
+
\int_{0}^{\infty} \mathrm{d} \lambda \frac{4(\lambda^{2}-m_{l}^{2}) \partial \rho / \partial m_{l}}{(\lambda^{2}+m_{l}^{2})^{2}}
~.
\end{equation}
Here we define the first part of $\partial \chi_{\text {disc }}/\partial m_{l}$, which is related to $\partial^2\rho(\lambda , m_l)/\partial m_l^2$, as $\chi_2$.

\begin{figure}[ht!]
\includegraphics[width=0.33\textwidth]{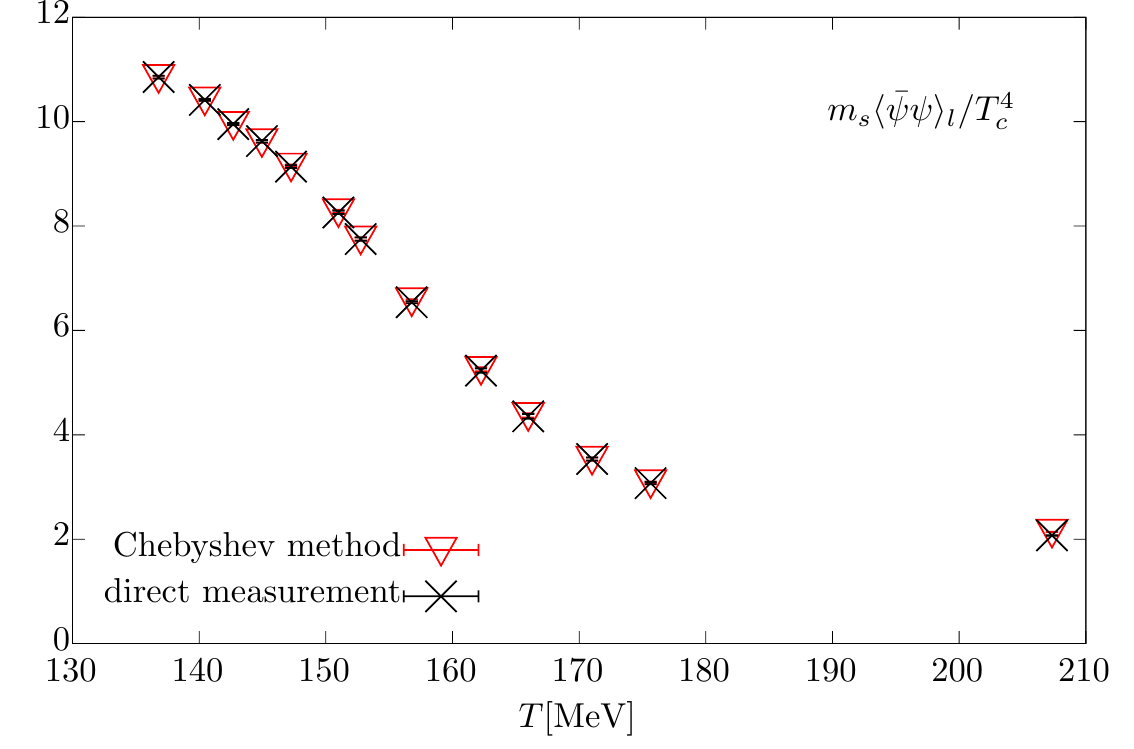}
\includegraphics[width=0.33\textwidth]{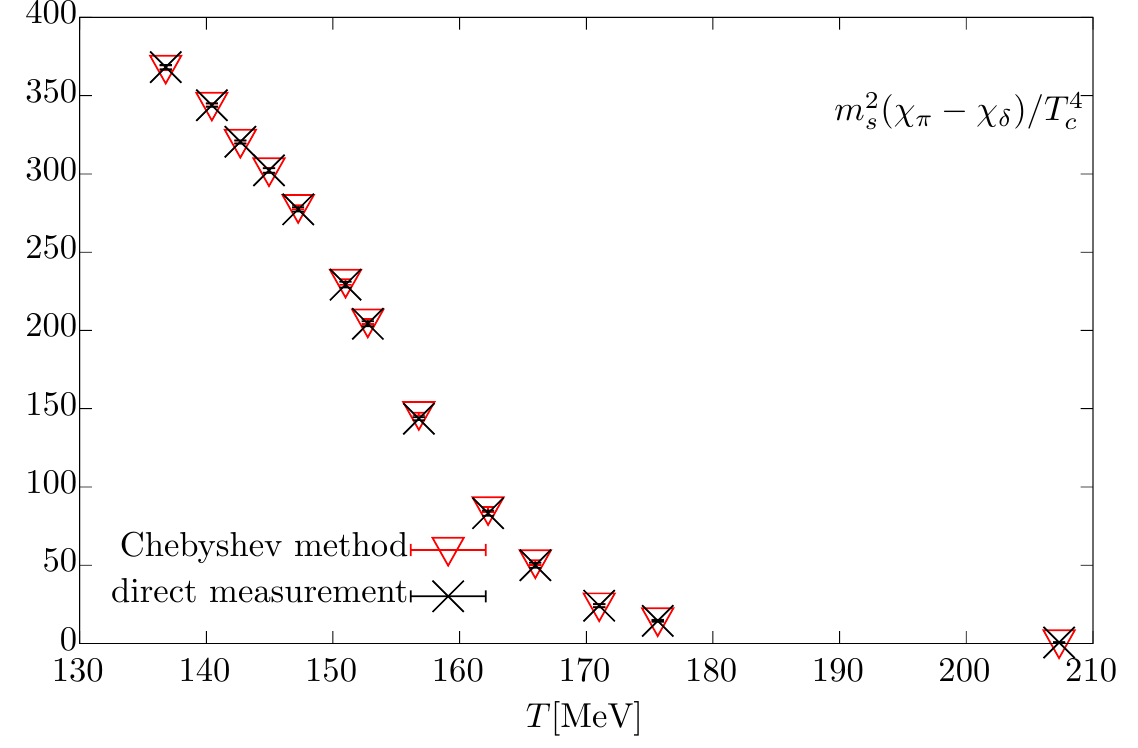}
\includegraphics[width=0.33\textwidth]{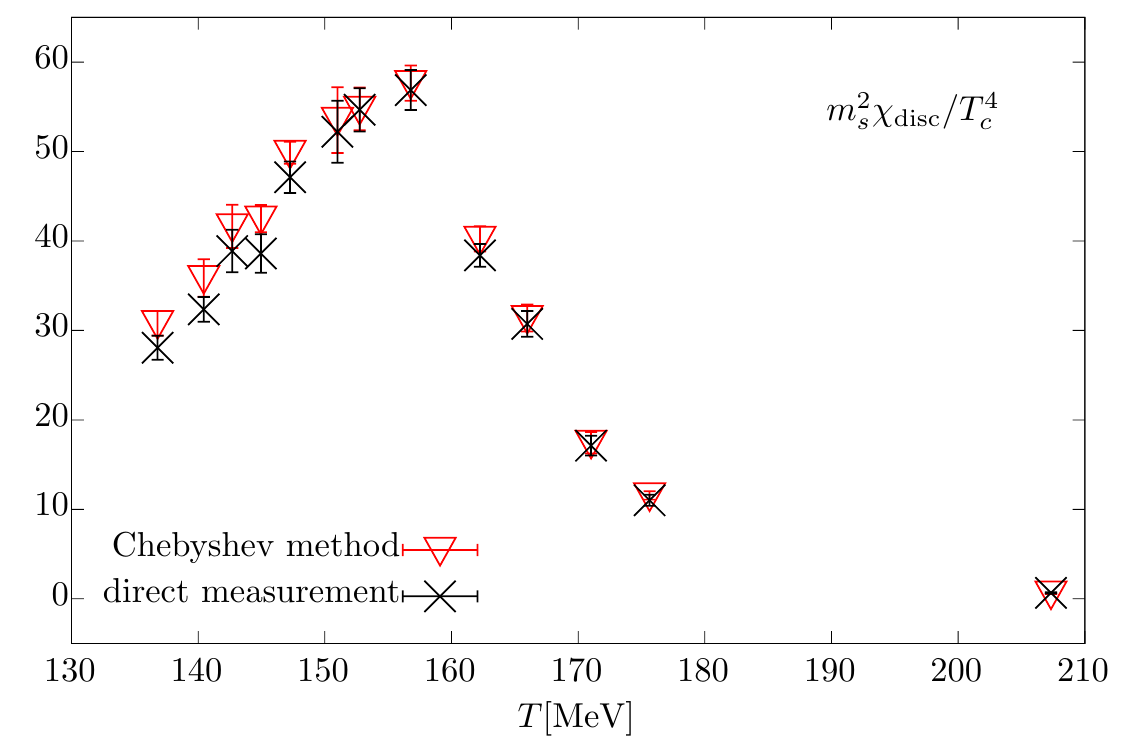}
\caption{Comparisons of direct measurements (black points) of the light quark chiral condensate $\langle\bar{\psi}\psi\rangle_l=\frac{2 T}{V} \langle \operatorname{Tr} M_l^{-1}\rangle$ (left), $\chi_{\pi}-\chi_{\delta}=\frac{2 T}{V} \frac{\langle \operatorname{Tr} M_l^{-1}\rangle}{m_l} + \frac{2 T}{V} \langle \operatorname{Tr} M_l^{-2}\rangle$ (middle) and the disconnected susceptibility $\chi_{\mathrm{disc}}=\frac{4 T}{V} [\langle(\operatorname{Tr} M_{l}^{-1})^{2}\rangle-\langle\operatorname{Tr} M_{l}^{-1}\rangle^{2}]$ (right) with those reproduced from $\rho$ (cf. Eq.(\ref{pbp_l}), (\ref{chi_pi-chi_delta})) and $\partial \rho/\partial m_l$ (cf. Eq.(\ref{chi_disc})) respectively (red points).}
\label{comparison_pbp_chidisc}
\end{figure}

\begin{figure}[htbp]
\includegraphics[width=.5\textwidth]{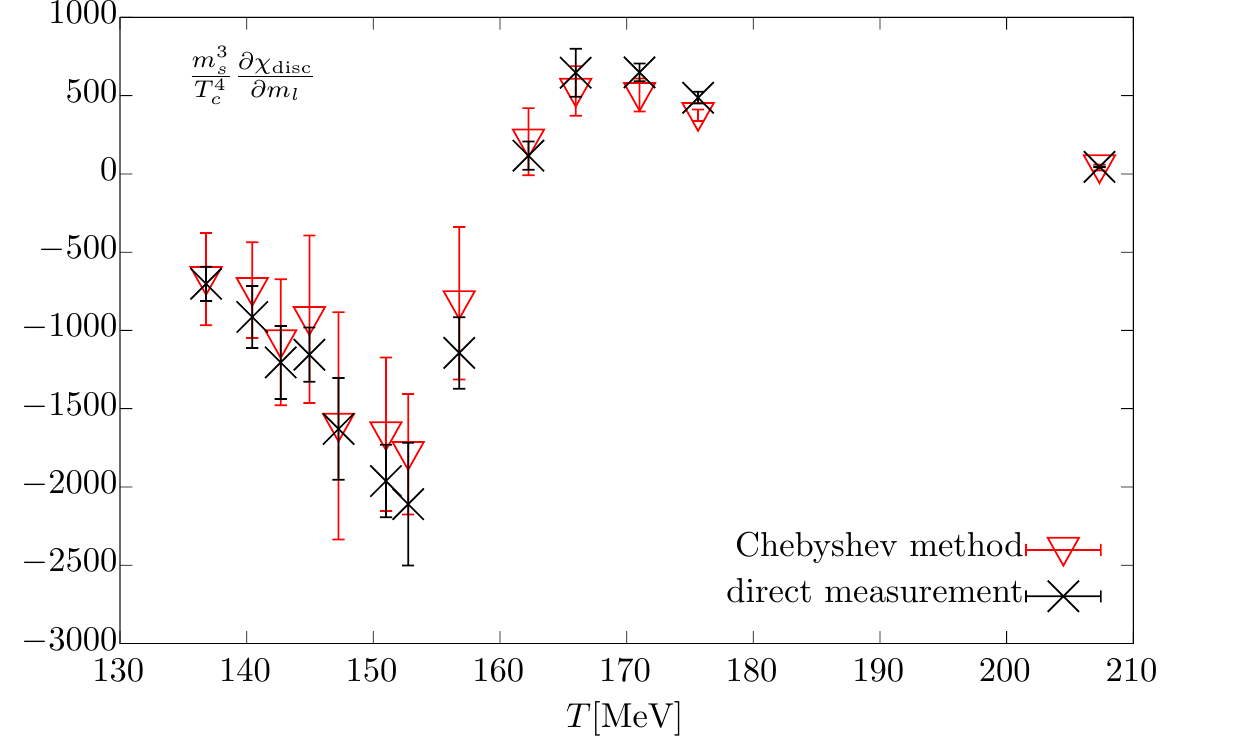}
\includegraphics[width=.5\textwidth]{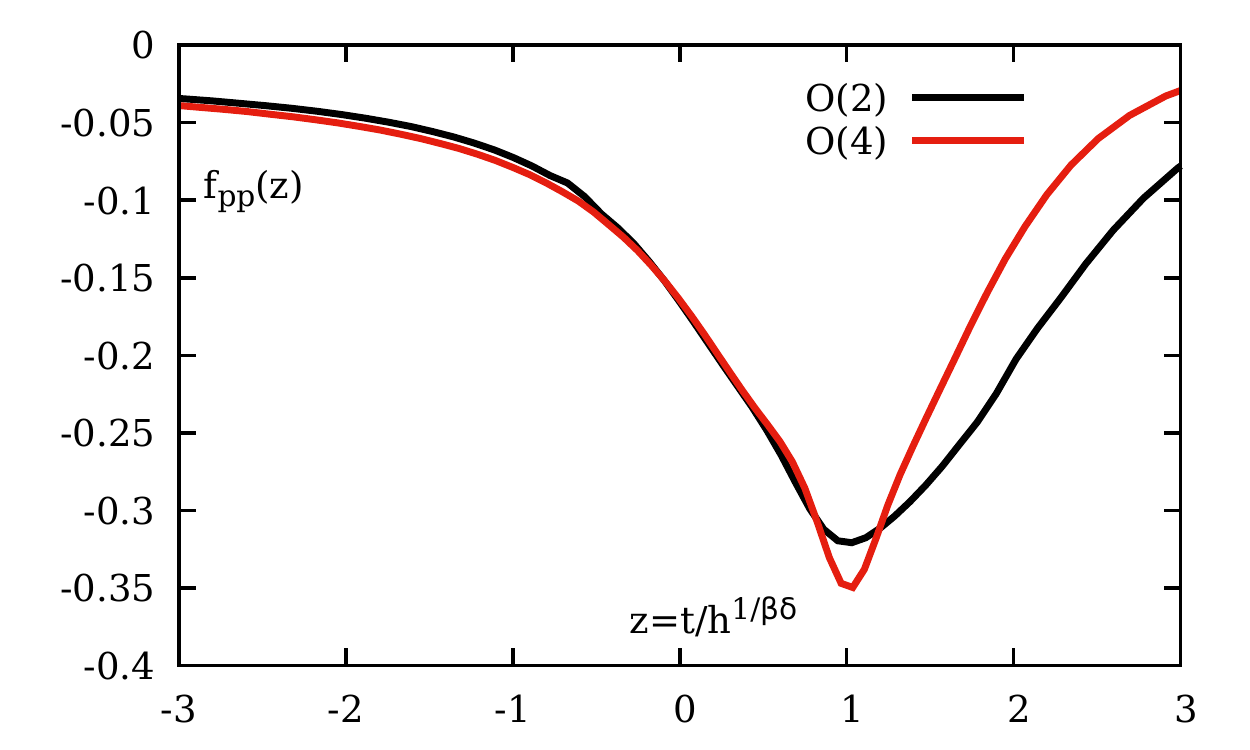}
\caption{Left: The mass derivative of the disconnected susceptibility, $\partial \chi_{\mathrm{disc}}/\partial m_l$, given by direct measurement $\partial \chi_{\mathrm{disc}}/\partial m_l = \frac{T}{V} 
[-24 \langle\operatorname{Tr} M_{l}^{-1}\rangle \langle(\operatorname{Tr} M_{l}^{-1})^{2}\rangle
+ 8 \langle(\operatorname{Tr} M_{l}^{-1})^{3}\rangle
- 8 \langle \operatorname{Tr} M_{l}^{-1}\operatorname{Tr} M_{l}^{-2}\rangle
+ 16\langle \operatorname{Tr} M_{l}^{-1}\rangle^{3}
+ 8 \langle \operatorname{Tr} M_{l}^{-1}\rangle \langle \operatorname{Tr} M_{l}^{-2}\rangle]$ (black points) and reproduced by $\partial \rho/ \partial m_l$ and $\partial ^2\rho/ \partial m_l^2$ (cf. Eq.(\ref{mass_chi_rho})) (red points). Right: The scaling function $f_{\mathrm{pp}}=(1 / \delta-1) f_{G}(z)+\left(\frac{z}{\beta}+\frac{z}{\beta^{2} \delta}-2 \cdot \frac{z}{\beta \delta}\right) f_{G}^{\prime}(z)+\frac{z^{2}}{\beta^{2} \delta} f_{G}^{\prime \prime}(z)$ for $O(2)$ (black line) and $O(4)$ (red line) universality class.}
\label{mass_derivative_chi_disc}
\end{figure}

When computing $\rho(\lambda,m_l)$  with Chebyshev filtering method the bin size in $\lambda$ needs to be fixed. To do so we choose the bin size in $\lambda$ such that $\chi_{\pi}-\chi_{\delta}$ given by Eq.(\ref{chi_pi-chi_delta}) can reproduce directly measured result. Just as confirmed in Ref.\cite{Ding:2020xlj}, once the bin size in $\lambda$ in the numerical integration of Eq.(\ref{chi_pi-chi_delta}) is fixed to reproduce directly measured $\chi_{\pi}-\chi_{\delta}$, the same bin size can also be used to reproduce $\langle \bar{\psi}\psi\rangle_l$, $\chi_{\mathrm{disc}}$ and $\partial \chi_{\mathrm{disc}} / \partial m_l$.
Comparisons of  $\langle \bar{\psi}\psi\rangle_l$, $\chi_{\pi}-\chi_{\delta}$, $\chi_{\mathrm{disc}}$ and $\partial \chi_{\mathrm{disc}}/\partial m_l$ computed by direct measurement with those reproduced from $\rho$ and its mass derivatives are shown in Fig.\ref{comparison_pbp_chidisc} and the left plot of Fig.\ref{mass_derivative_chi_disc}. The consistency of these observables given by the two different methods shows that the computations of $\rho$ and $\partial^n\rho(\lambda, m_l)/\partial m_l^n (n=1,2)$ are reliable.

In left plot of Fig.~\ref{mass_derivative_chi_disc} we show the temperature dependence of $\partial \chi_{\mathrm{disc}}/\partial m_l$. It can be clearly observed that as $T$ increases $\partial \chi_{\mathrm{disc}}/\partial m_l$ firstly decreases at $T\lesssim 155$ MeV, then increase till a turning point at $T\approx 165$ MeV and finally approaches 0 at $T\approx 205$ MeV. It is also interesting to see that $\partial \chi_{\mathrm{disc}}/\partial m_l$ is negative at $T\lesssim 165$ MeV and flip its sign at $T\gtrsim 165$ MeV. The temperature dependence of  $\partial \chi_{\mathrm{disc}}/\partial m_l$ thus is consistent with that of the scaling function $f_{\mathrm{pp}}$ as shown in the right plot of Fig.~\ref{mass_derivative_chi_disc}.

\begin{figure}[htbp]
\includegraphics[width=.48\textwidth]{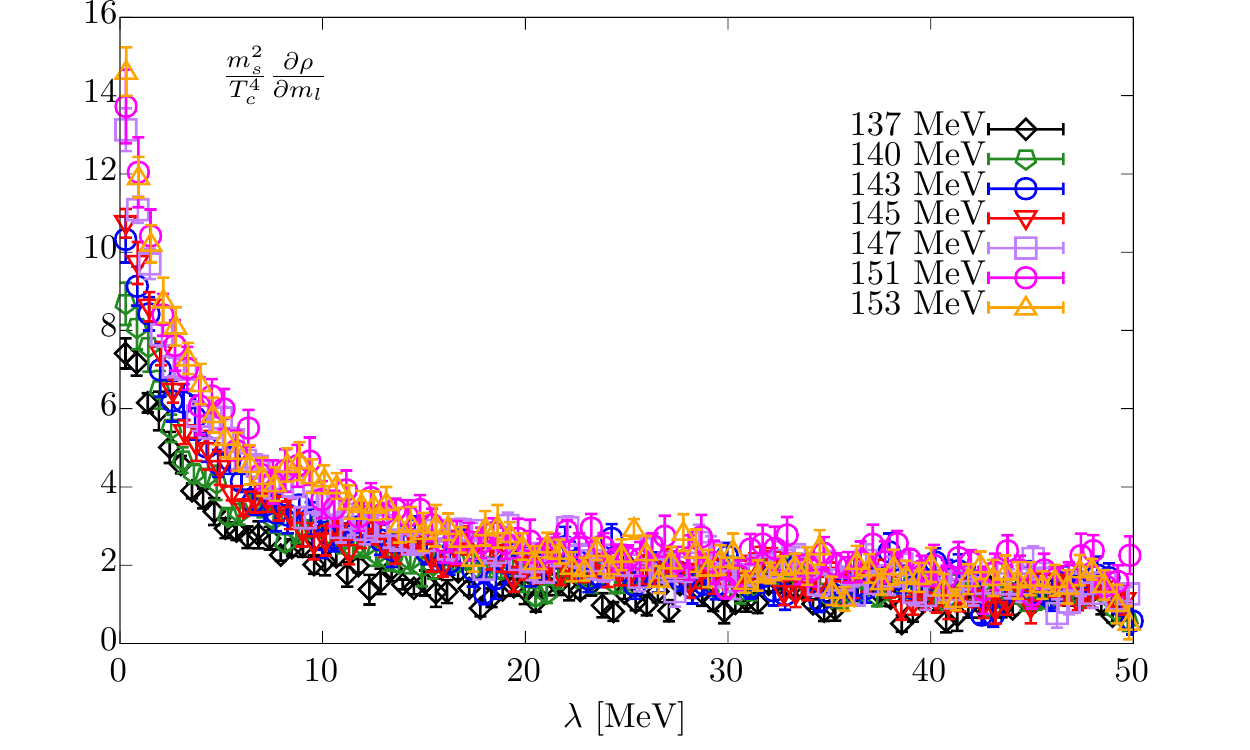}
\includegraphics[width=.48\textwidth]{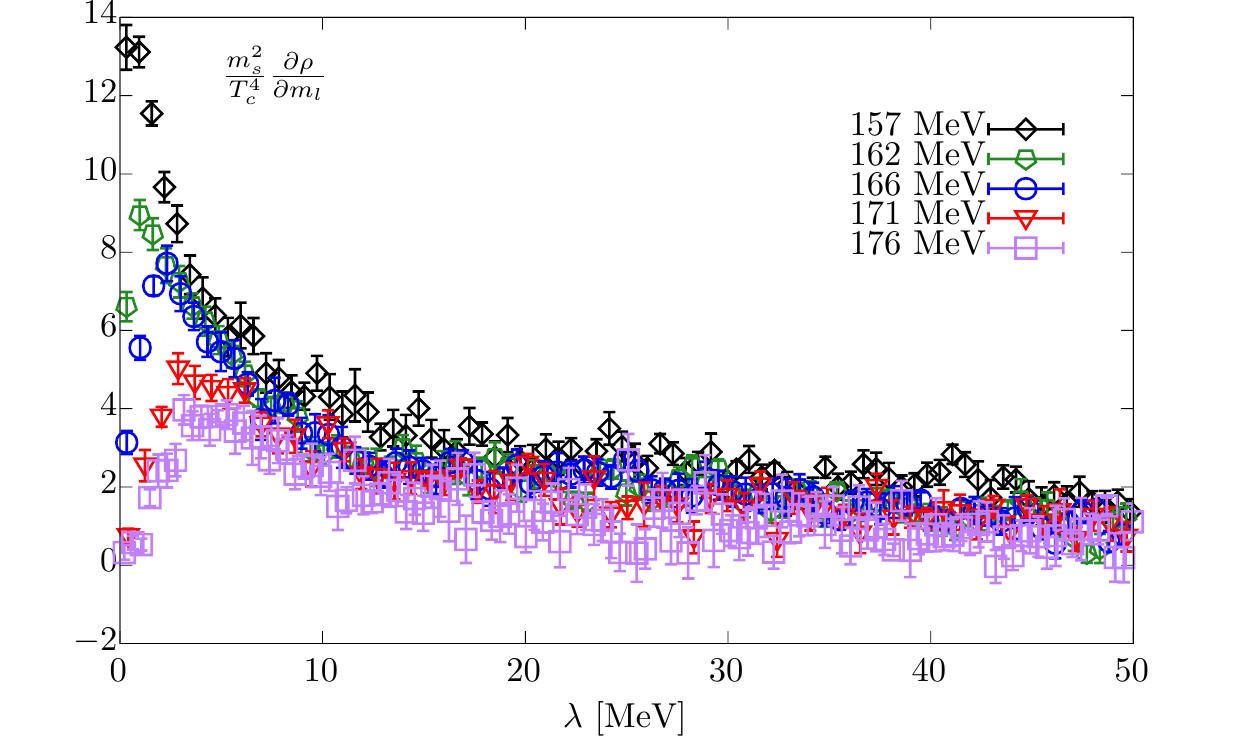}
\caption{$\partial \rho / \partial m_{l}$ for temperatures below 153 MeV (left) and above 157 MeV (right).}
\label{par1}
\end{figure}

To study the microscopic origin of the scaling behaviors of $\chi_{\mathrm{disc}}$ and $\partial \chi_{\mathrm{disc}}/\partial m_l$, we now directly investigate $\partial \rho/ \partial m_l$ and $\partial^2 \rho/ \partial m_l^2$. We show the results of $\partial \rho(\lambda, m_l)/\partial m_l$ at $T\lesssim153$ MeV and $T\gtrsim157$ MeV in the left and right plots of Fig.\ref{par1}, respectively.
At temperature $T$ below 153 MeV, the amplitude of $\partial \rho/\partial m_l$ in the small $\lambda$ region increases with $T$ monotonously, while at $T$ above 157 MeV the amplitude decreases with $T$ instead.
Since the infrared part (small $\lambda$ region) of $\partial \rho/\partial m_l$ dominates the contribution to $\chi_{\mathrm{disc}}$, this non-monotonous temperature dependence of $\partial \rho/\partial m_l$ in the infrared region is consistent with that of $\chi_{\mathrm{disc}}$ as shown in the right plot of Fig.\ref{comparison_pbp_chidisc}.

\begin{figure}[htbp]
\includegraphics[width=.5\textwidth]{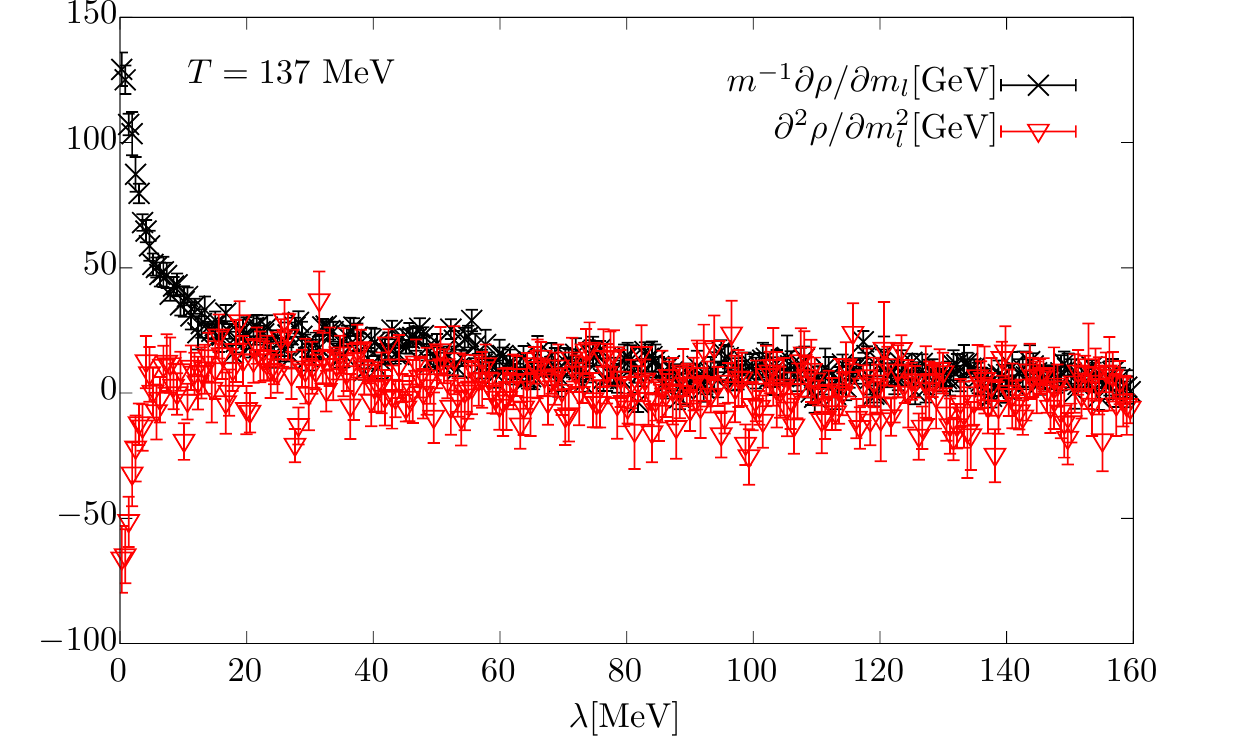}
\includegraphics[width=.5\textwidth]{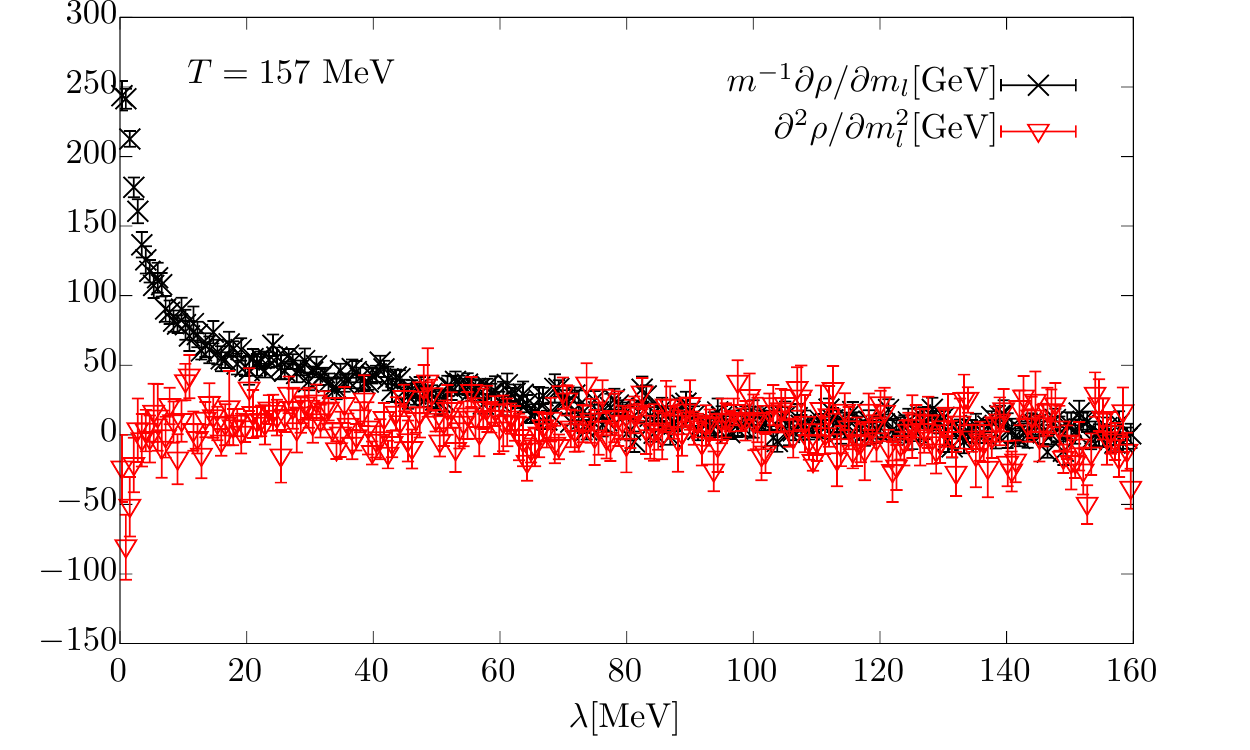}
\includegraphics[width=.5\textwidth]{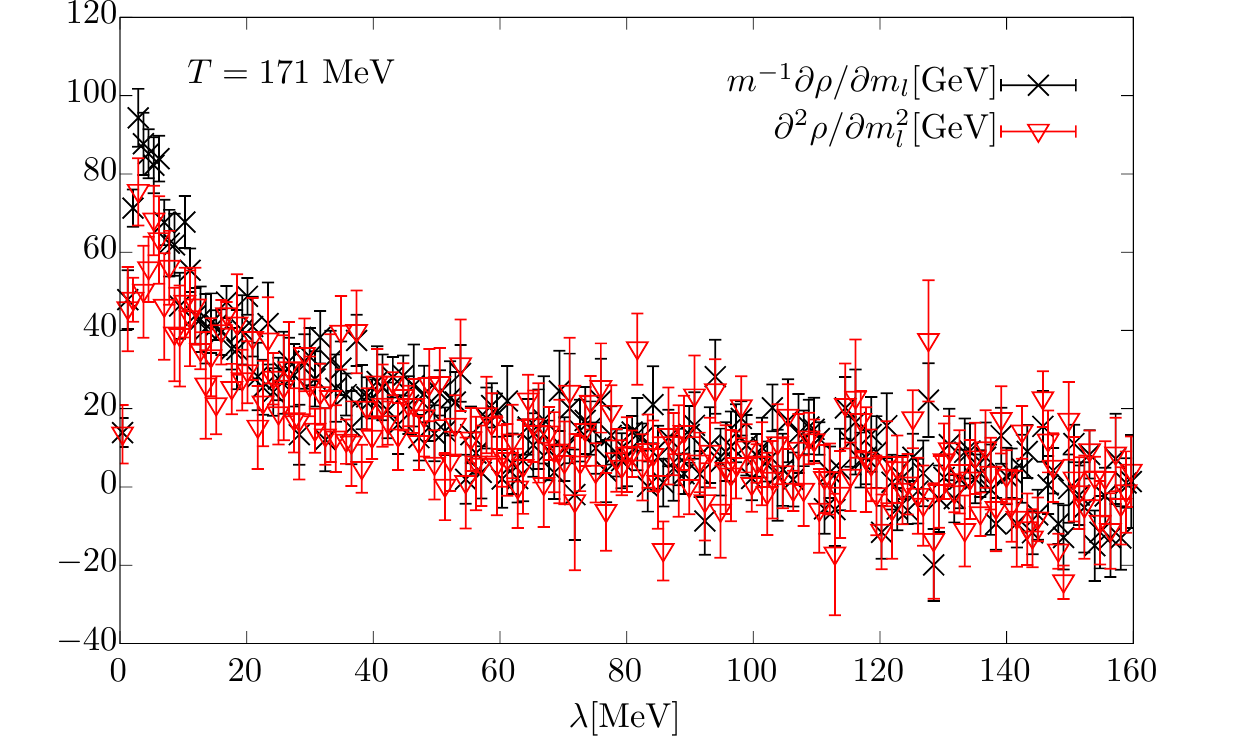}
\includegraphics[width=.5\textwidth]{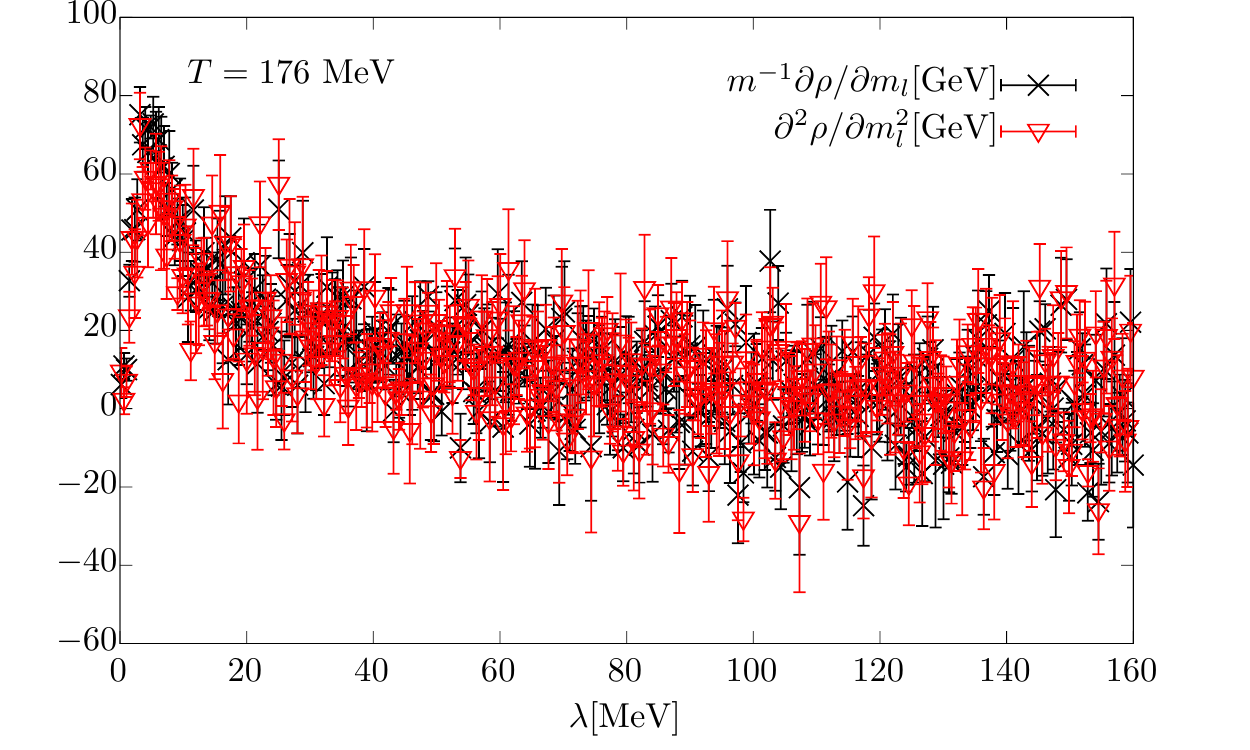}
\caption{Comparisons of $m_{l}^{-1} \partial \rho / \partial m_{l}$ (black points) with $\partial^{2} \rho / \partial m_{l}^{2}$ (red points) for T=137 MeV (top left), 157 MeV (top right), 171 MeV (bottom left) and 176 MeV (bottom right).}
\label{par2}
\end{figure}

At a single temperature $T=1.6 T_c\approx 205$ MeV it was observed in Ref.~\cite{Ding:2020xlj} that $m_{l}^{-1} \partial \rho / \partial m_{l} \approx \partial^{2} \rho / \partial m_{l}^{2}$ and $\partial^{3} \rho / \partial m_{1}^{3} \approx 0$. This leads to $\rho\propto m_l^2$ consistent with the dilute instanton gas approximation at $T\simeq 1.6 T_c$ MeV.
In Fig.\ref{par2} we now confront $m_{l}^{-1} \partial \rho / \partial m_{l}$ with $\partial^{2} \rho / \partial m_{l}^{2}$ at temperatures much lower than 1.6 $T_c$, i.e. ranging from 137 MeV to 176 MeV. It can be observed that $m_{l}^{-1} \partial \rho / \partial m_{l}$ and $\partial^{2} \rho / \partial m_{l}^{2}$ are no longer consistent with each other for temperatures from 137 MeV to 171 MeV, and $\partial^{2} \rho / \partial m_{l}^{2}$ in the infrared region even becomes negative at temperatures less than 157 MeV.

Assuming $\rho\propto m_l^c$ we will have $m_{l}^{-1} \partial \rho / \partial m_{l}\propto cm_l^{c-2}$ and $\partial^{2} \rho / \partial m_{l}^{2}\propto c(c-1)m_l^{c-2}$. 
Then the assumption $c=2$ leads to $m_{l}^{-1} \partial \rho / \partial m_{l} = \partial^{2} \rho / \partial m_{l}^{2}$.
Fig.\ref{par2} tells us $m_{l}^{-1} \partial \rho / \partial m_{l} \neq \partial^{2} \rho / \partial m_{l}^{2}$ for temperatures around $T_c$, which suggests $c \neq 2$ for this temperature region from 137 MeV to 171 MeV. This means that as temperature approaches to $T_c$  the spectral density $\rho$ is no longer proportional to $m_l^2$.
This in turn implies that the dilute instanton gas approximation is not valid in the vicinity of $T_c$.
Based on the fact that $\partial \rho / \partial m_{l}$ is always positive we conclude that $c>0$ in the current tempearture window. 
On the other hand, in the vicinity of $T_c$ $\partial^{2} \rho / \partial m_{l}^{2} <0$ around $T_c$ suggests $c<1$. Thus given $\rho \propto m_l^c$ the power $c$ is in the range of $(0,1)$ around $T_c$.

\section{Scaling in correlations of Dirac eigenvalues}

At temperature around $T_c$, scaling behaviors in temperature have been observed in the mass derivatives of Dirac eigenvalue spectrum, i.e. in the correlation of the Dirac Eigenvalues (cf. Figs.~\ref{par1} and~\ref{par2}). Inspired by the Dirac eigenvalue spectrum form of $\chi_{\mathrm{disc}}$ (Eq.(\ref{chi_disc})) and $\chi_2$ (Eq.(\ref{mass_chi_rho})), we may expect that $\partial \rho(T;\lambda, m_l)/\partial m_l$ and $\partial^2 \rho(T;\lambda, m_l)/ \partial m_l^2$ can be factorized into two parts
\begin{align}
\label{par1_factorize}
\frac{\partial \rho(T;\lambda, m_l)}{\partial m_l}&= f_1(T)\cdot g_1(\lambda,m_l)~,
\\
\label{par2_factorize}
\frac{\partial^2 \rho(T;\lambda, m_l)}{\partial m_l^2}&= f_2(T)\cdot g_2(\lambda,m_l)~,
\end{align}
where the temperature dependence is only encoded in $f_n(T)~(n=1,2)$. Then we can constructed the ratio $R_n(T)~(n=1,2)$ defined as
\begin{align}
R_1(T)
\equiv \frac{\chi_{\mathrm{disc}}(T)}{\chi_{\mathrm{disc}}(T_0)}
=\frac{ \int_{0}^{\infty} \mathrm{d} \lambda \frac{4 m_{l}  \cdot { \partial \rho(T;\lambda, m_l) / \partial m_{l}}}{\lambda^{2}+m_{l}^{2}} }{ \int_{0}^{\infty} \mathrm{d} \lambda \frac{4 m_{l}  \cdot { \partial \rho(T_0;\lambda, m_l) / \partial m_{l}}}{\lambda^{2}+m_{l}^{2}} }
=\frac{ \int_{0}^{\infty} \mathrm{d} \lambda \frac{4 m_{l}  \cdot  { f_1(T)\cdot g_1(\lambda,m_l)} }{\lambda^{2}+m_{l}^{2}} }{ \int_{0}^{\infty} \mathrm{d} \lambda \frac{4 m_{l}  \cdot  { f_1(T_0)\cdot g_1(\lambda,m_l)} }{\lambda^{2}+m_{l}^{2}} }
=\frac{f_1(T)}{f_1(T_0)}
~,
\end{align}
\begin{align}
R_2(T) 
\equiv \frac{\chi_2(T)}{\chi_2(T_0)}
=\frac{ \int_{0}^{\infty} \mathrm{d} \lambda \frac{4 m_{l} \cdot { \partial^2 \rho(T;\lambda, m_l) / \partial m_{l}^2}}{\lambda^{2}+m_{l}^{2}}}{\int_{0}^{\infty} \mathrm{d} \lambda \frac{4 m_{l} \cdot { \partial^2 \rho (T_0;\lambda, m_l)/ \partial m_{l}^2}}{\lambda^{2}+m_{l}^{2}}}
=\frac{\int_{0}^{\infty} \mathrm{d} \lambda \frac{4 m_{l}  \cdot { f_2(T)\cdot g_2(\lambda,m_l) }}{\lambda^{2}+m_{l}^{2}}}{\int_{0}^{\infty} \mathrm{d} \lambda \frac{4 m_{l}  \cdot { f_2(T_0)\cdot g_2(\lambda,m_l) }}{\lambda^{2}+m_{l}^{2}}}
=\frac{f_2(T)}{f_2(T_0)} ~,
\end{align}
where $T_0$ is an arbitrary temperature chosen as a reference value (the lowest temperature is set as $T_0$ in this work, i.e. $T_0$=137 MeV). With the help of $R_n(T)$ the temperature dependence in $\partial \rho(\lambda, m_l)/\partial m_l$ and $\partial^2 \rho(\lambda, m_l)/ \partial m_l^2$ can be removed in the following way
\begin{align}
\label{eq:Tfactorization1}
\frac{\partial \rho/\partial m_l}{ R_1(T)} 
&=\frac{f_1(T)}{R_1(T)}\times  g_1(\lambda,m_l)
=f_1(T_0)\times  g_1(\lambda,m_l)
~,
\\
\label{eq:Tfactorization2}
\frac{\partial^2 \rho/\partial m_l^2}{ R_2(T)} 
&=\frac{f_2(T)}{R_2(T)} \times g_2(\lambda,m_l)
=
f_2(T_0)\times
g_2(\lambda,m_l)~.
\end{align}
Thus the temperature independence of ${\partial \rho/\partial m_l} / { R_1(T)}$ would indicate that the factorization in Eq.(\ref{par1_factorize}) works. The similar argument holds for $\partial^2 \rho / \partial m_l^2/R_2(T)$ and Eq.(\ref{par2_factorize}).

\begin{figure}[htbp]
\includegraphics[width=.48\textwidth]{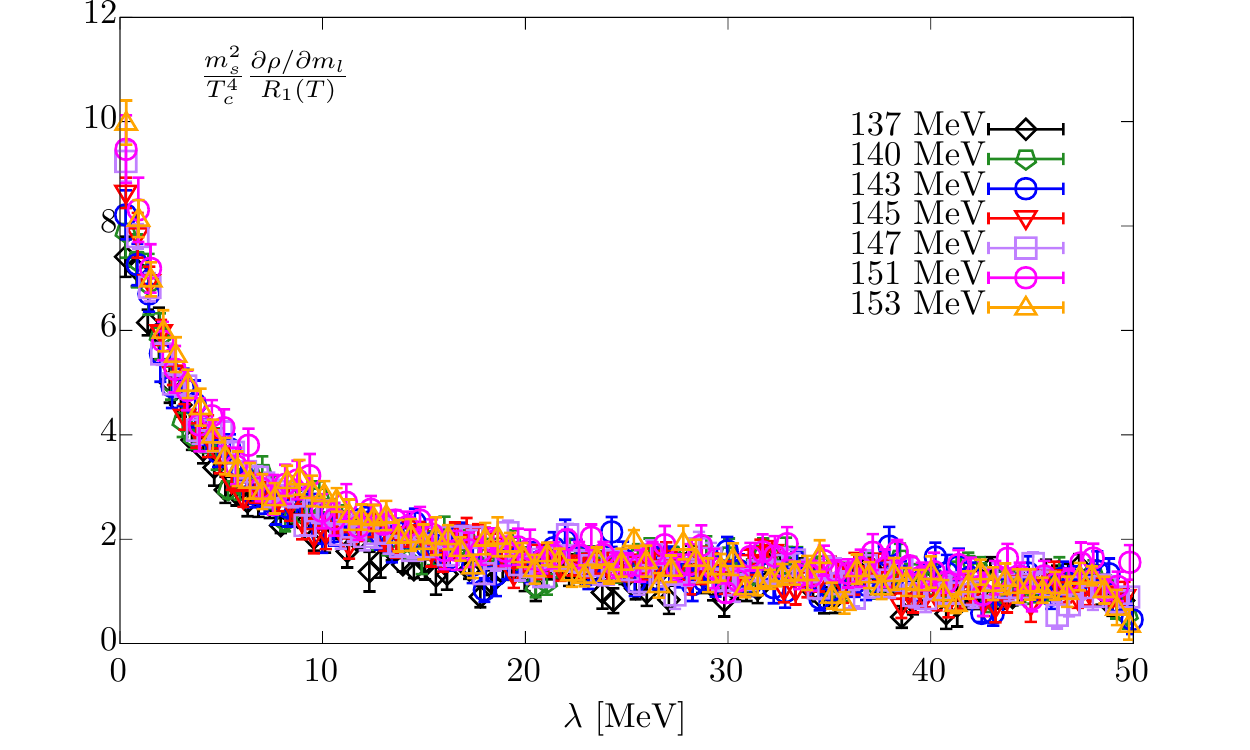}
\includegraphics[width=.48\textwidth]{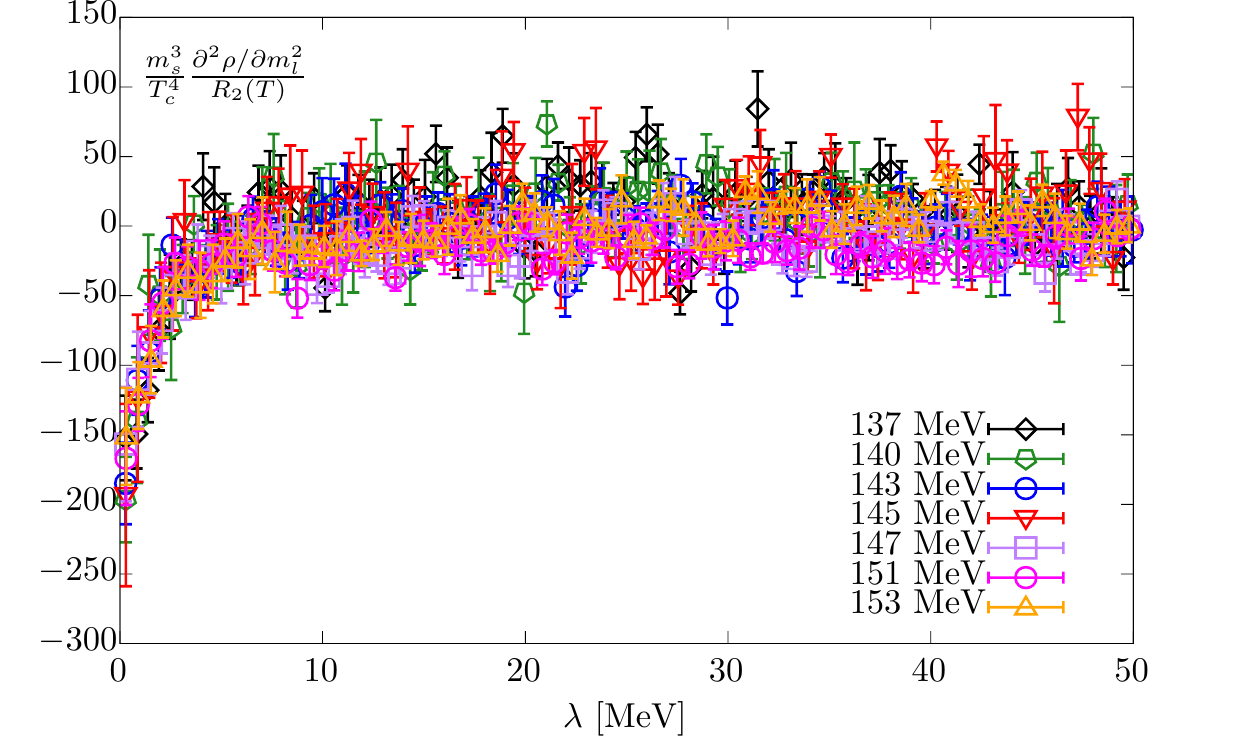}
\caption{${\partial \rho/\partial m_l} / { R_1(T)}$ (left) and ${\partial^2 \rho/\partial m_l^2} / { R_2(T)}$ (right) at temperatures from 137 MeV to 153 MeV.}
\label{fig_par1_factorize}
\end{figure}

We show ${\partial \rho/\partial m_l} / { R_1(T)}$ and ${\partial^2 \rho/\partial m_l^2} / { R_2(T)}$ at temperatures around $T_c$ in the left and right plots of Fig.\ref{fig_par1_factorize}, respectively. It can be observed that ${\partial \rho/\partial m_l} / { R_1(T)}$ at temperatures from 137 MeV to 153 MeV are almost temperature independent. This indicates that the temperature dependence in $\partial \rho / \partial m_l$ can be indeed factored out in the critical region. The similar conclusion can be drawn for $\partial^2 \rho / \partial m_l^2$.

Based on the factorization of ${\partial \rho/\partial m_l}$ and ${\partial^2 \rho/\partial m_l^2}$  shown in Eq.(\ref{par1_factorize}) and Eq.(\ref{par2_factorize}) it is natural to expect that $\rho$ can also be factorized as $f_0(T) \times g_0(\lambda,m_l )$. Considering that $g_0(\lambda,m_l )$ is proportional to $m^c_l$, we have
\begin{align}
\rho =f_0(T)\times g_0(\lambda,m_l)
 \propto f_0(T) \times {  m^c_l}
 ~.
\end{align}
Thus Eq.(\ref{eq:Tfactorization1}) and Eq.(\ref{eq:Tfactorization2}) can be further expressed as follows
\begin{align}
\frac{\partial \rho/\partial m_l}{ R_1(T)} 
&=f_1(T_0)\times  g_1(\lambda,m_l)
\propto{ c \cdot m_l^{c-1}}~,
\\
\frac{\partial^2 \rho/\partial m_l^2}{ R_2(T)} 
&=
f_2(T_0)\times
g_2(\lambda,m_l)\propto
{  c\cdot (c-1) m_l^{c-2} }
~.
\end{align}
Hence $c$ can be further estimated by comparing
$\partial \rho/\partial m_l/R_1(T)$ and $m_l/(c-1) \times \partial^2 \rho/\partial m_l^2/R_2(T)$.

\begin{figure}[H]
\centering
\includegraphics[width=.6\textwidth]{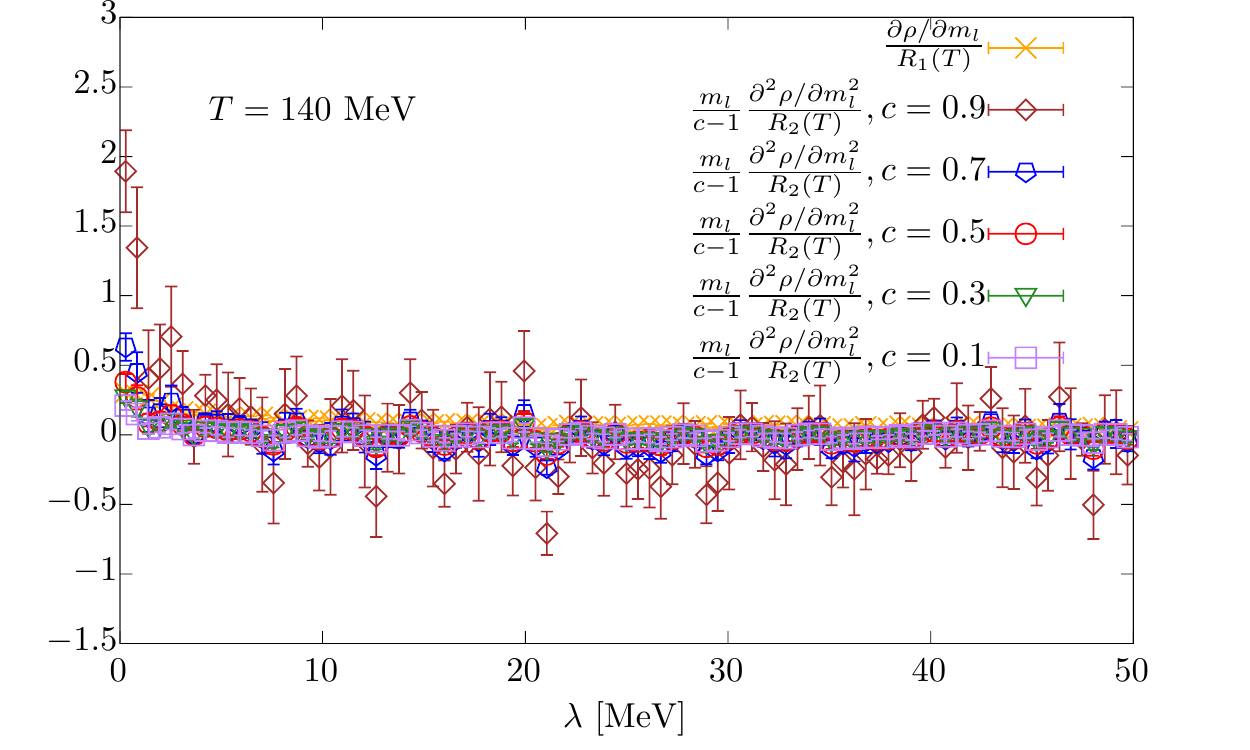}
\caption{Comparison of $\partial \rho/\partial m_l/R_1(T)$ (cross symbol) with $m_l/(c-1) \times \partial^2 \rho/\partial m_l^2/R_2(T)$ with different values of $c$ (open symbols in different colors) at $T=140$ MeV.}
\label{estimate_c}
\end{figure}

Fig.\ref{estimate_c} shows the comparison of $\partial \rho/\partial m_l/R_1(T)$ with $m_l/(c-1) \times \partial^2 \rho/\partial m_l^2/R_2(T)$ with different values of $c$ at $T=140$ MeV. 
We can observe that as $c$ decreases from 0.9 to 0.3, $\partial \rho/\partial m_l/R_1(T)$ and $m_l/(c-1) \times \partial^2 \rho/\partial m_l^2/R_2(T)$ seem to agree with each other better and better; then as $c$ decreases further from 0.3 to 0.1 they start to deviate from each other. Hence we can further restrict the range of $c$ from $(0,1)$ to $(0.1,0.7)$ for $T=140$ MeV.

\section{Conclusions}

In this work we investigate the criticality manifested in $\partial \rho(\lambda, m_l ) / \partial m_l$ and $\partial^2 \rho(\lambda, m_l ) / \partial m_l^2$ in (2+1)-flavor QCD in the temperature window $T\in[137,176]$ MeV.
The (2+1)-flavor lattice QCD simulations are performed using HISQ fermions on $N_\tau=8$ lattices with $m_\pi=110$ MeV.

We have found that $m_l^{-1}\partial \rho/\partial m_l$ is no longer equal to $\partial ^2\rho/\partial m_l^2$ in the current temperature window, which is shown a quite different behavior compared with that at high temperature about $1.6T_c$. And $\partial ^2\rho/\partial m^2$ in the infrared region even becomes negative at certain low temperatures. These suggest that as temperature approaches to $T_c$ the $m_l^2$ behavior in $\rho$ does not exist any more, and dilute instanton gas approximation is not valid in this temperature window. 
Based on the assumption of $\rho\propto m_l^c$, the range of the power $c$ should be in $(0,1)$ for temperatures around the transition temperature. 
We further demonstrated that at $T \in [137, 153]$ MeV it seems that the temperature dependence can be factored out in $\partial \rho/ \partial m_l$ and $\partial^2 \rho/ \partial m_l^2$. Consequently the value of $c$ can be pin down by comparing ${(\partial \rho/\partial m_l)}/{ R_1(T)}$ and $m_l/(c-1) \times {(\partial^2 \rho/\partial m_l^2)}/{ R_2(T)}$. 

In our current study only one single pion mass is used. In the near future we will perform detailed studies of the spectral
density in (2+1)-flavor QCD with other quark masses.

\acknowledgments
This material is based upon work supported by the National Natural Science Foundation of China under the grant number 11775096 and the U.S. Department of Energy, Office of Science, Office of Nuclear Physics through the Contract No. DE-SC0012704 and within the framework of Scientific Discovery through Advance Computing (SciDAC) award "Computing the Properties of Matter with Leadership Computing Resources".
The numerical simulations have been performed on the GPU cluster in the Nuclear Science Computing Center at Central China Normal University (NSC$^3$), Wuhan, China and the facilities of the USQCD Collaboration, which are funded by the Office of Science of the U.S. Department of Energy.
For generating the gauge configurations, the HotQCD software suite was used, and the eigenvalue measurement code was developed also based on the same software suite.

\bibliographystyle{JHEP.bst}
\bibliography{cited.bib}



\end{document}